\documentclass{article}
\usepackage{amsmath}
\usepackage{epsfig}
\begin{document}
\begin{titlepage}

\begin{flushright}
USITP-00-16\\
hep-th/0010083 
\end{flushright}

\vspace{1cm}

\begin{center}
{\huge\bf  Multi-Black Hole Sectors \\ 
\vspace{3mm} 
of AdS\(_{3}\) Gravity}
\end{center}
\vspace{5mm}

\begin{center}
    
{\large Teresia M\aa nsson and Bo Sundborg}\\

\vspace{3mm}

Institute of Theoretical Physics, Box 6730, SE-113 85 
Stockholm, Sweden

\vspace{3mm}

{\tt
teresia@physto.se, bo@physto.se\\
}

\end{center}

\vspace{5mm}

\begin{center}
{\large \bf Abstract}
\end{center}
\noindent We construct and discuss solutions of \(SO(1,2)\times
SO(1,2)\) Chern-Simons theory which correspond to multiple BTZ black
holes.  These solutions typically have additional singularities, the
simplest cases being special conical singularities with a \(2\pi\)
surplus angle.  There are solutions with singularities inside a common
outer horizon, and other solutions with naked conical singularities. 
Previously such singularities have been ruled out on physical
grounds, because they do not obey the geodesic equation.  We find
however that the Chern-Simons gauge symmetry may be used to locate all
such singularities to the horizons, where they necessarily follow
geodesics.  We are therefore led to conclude that these singular
solutions correspond to physically sensible geometries.

Boundary charges at infinity are only sensitive to the total mass and
spin of the black holes, and not to the distribution among the black
holes.  We therefore argue that a holographic description in terms of
a boundary conformal field theory should represent both single and
multiple BTZ solutions with the same asymptotic charges.  Then sectors
with multiple black holes would contribute to the black hole entropy
calculated from a boundary CFT.

\vfill
\begin{flushleft}
October 2000
\end{flushleft}
\end{titlepage}
\newpage

\newpage
 
\section{Introduction}
Three-dimensional gravity has been a useful laboratory for exploring
quantum gravity in a simplified setting.  For a negative cosmological
constant there are black hole solutions
\cite{Banados:1992wn,Banados:1993gq}, and the Bekenstein-Hawking
entropy of these BTZ black holes have been attributed to boundary
degrees of freedom at the horizon \cite{Carlip:1995gy,Banados:1997ad},
at infinity \cite{Strominger:1998eq} or at any intermediate timelike
surfaces \cite{Banados:1999ta}.

Strominger's asymptotic approach makes use of a particular property of
asymptotically AdS solutions of three-dimensional gravity discovered
by Brown and Henneaux \cite{Brown:1986nw}: the asymptotic isometries
are represented canonically by a Virasoro algebra.  The BTZ mass and
spin determine the transformation properties under the conformal
transformations (conformal weights).  In conformal field theory an
argument by Cardy \cite{Cardy:1986ie} can be used to relate central
charges of Virasoro algebras to the densities of states at high
weights.  Similarly the asymptotic density of states of quantum
gravity is fixed by the central charge.  It is found to agree with
expectations from the BTZ horizon area.  This elegant argument is 
independent of the precise conformal field theory representing quantum
gravity, and gives few details about the theory.  Carlip has combined
it with his horizon approach, to open the way to an understanding of
the universal nature of black hole entropy \cite{Carlip:1999wz}, and
its relation to horizon area.  The price is that the horizon is
treated as an input, rather than as a consequence of the global
geometry, and again that details of the field theory are lost.

The ability to compute black hole entropy quantum mechanically does
not mean that it is fully explained.  Even in the simple case of 2+1
dimensions the gravitational backgrounds that contribute to the
entropy are poorly understood.  Ideally a correct count of boundary
degrees of freedom at the horizon or at infinity should also tell us
what bulk geometries are relevant, and how they are excited.  They may
also be represented differently in different quantum gravity theories. 
(In 2+1 dimensions there are several inequivalent quantizations
\cite{Carlip:1993zi}.)  To start investigating what geometries may
represent the entropy we choose to study one description, Chern-Simons
theory \cite{Achucarro:1986vz,Witten:1988hc}.  In Chern-Simon theory
the map to the boundary theory is well-known and produces a WZW theory
\cite{Witten:1989hf,Moore:1989yh,Elitzur:1989nr}.

On the classical level one could ask what constant curvature metrics
(solving the equations of motion) look asymptotically like BTZ black
holes, and could be expected to be equally important as the standard
BTZ solution for the black hole entropy.  Ba\~{n}ados
\cite{Banados:1998gg} (see also \cite{Ezawa:1995xm} and
\cite{Navarro-Salas:1998ks}) has given a simple analytic and general
characterization of such solutions, but unfortunately the analytic
expression of the solution does not give directly the geometric
structure of the spacetime.  In string theory approaches to black hole
entropy, BPS solutions which can be separated into multi-source
solutions play a prominent role \cite{Strominger:1996sh}.  This
indicates that similar solutions may be of interest also in pure
gravity.  Indeed, we will find that asymptotically, Chern-Simons
multi-source solutions typically are closer than most of the solutions
in \cite{Banados:1998gg} and \cite{Ezawa:1995xm} to the standard BTZ
solutions.

Chern-Simons multi-source solutions have been discussed as candidates
for stationary multi-black hole solutions by Coussaert and Henneaux
\cite{Coussaert:1994if}.  Clement \cite{Clement:1994qb} found similar
solutions in a metrical formulation and generalized them to dynamical
solutions with moving sources.  His main motivation for doing so was
to correct the shortcoming also observed in \cite{Coussaert:1994if},
that the stationary solutions necessarily involve additional conical
singularities which typically do not follow geodesics\footnote{The 
presence of these singularities indicates that such solutions are
different from the multi-black hole solutions with multiple asymptotic
regions that have been discussed by Brill
\cite{Brill:1998pr,Brill:2000xm} and the wormhole solutions by 
Bengtsson et al. \cite{Aminneborg:1998pz,Aminneborg:1999si}.}.  
A non-geodesic
behaviour signals an unwanted transport of energy-momentum between the
singularity and spacetime.  

From our perspective this state of affairs is quite puzzling.  The
multi-source solutions could be expected to be on equal footing with
the BTZ solutions on the basis of their asymptotic behaviour, and as
Chern-Simons solutions they are no less regular.  Hence calculations 
of black hole entropy based on Chern-Simons theory and its expression 
in terms of a boundary WZW model, which reproduce Bekenstein-Hawking's 
result, appear to include unphysical geometries.
Is Chern-Simons theory, which has attracted so much attention as a
model of quantum gravity, just not sensible as a theory of gravity?

Fortunately, there is a caveat in the above argument.  We find that
multi-source solutions have similar asymptotic behaviour, but also
that they represent physical (though singular) geometries.  The
crucial observation is that one should deal with gauge equivalence
classes of solutions to Chern-Simons theory rather than with
individual solutions.  As we show in section \ref{subsubsec.Geo}, all
``unacceptable" Coussaert-Henneaux solutions are gauge equivalent to
perfectly acceptable solutions\footnote{It may seem strange that some
gauge potentials give meaningful metrics while others in the same
class do not, but this property is in fact intrinsic to the
Chern-Simons approach.  The solutions to the equations of motion are
pure gauge, and they may locally be transformed away, giving a
completely degenerate metric, unless further conditions on the vector
potentials are imposed.}.  This is possible because the Chern-Simons
gauge group is larger than the diffeomorphism group.  For a discussion
of how the gauge symmetry conspires with the presence of degenerate
metrics in the Chern-Simons `formulation' see Matschull
\cite{Matschull:1999he}.  In practice, gauge transformations move the
static conical singularities to the horizons, where they obey the
geodesic equation (by infinite redshift) as already observed by
Clement.  If all Chern-Simons solutions have similar sensible
representative geometries or not is left as an open question.  The
Coussaert-Henneaux solutions dealt with here certainly constitute an
important subclass.

In section \ref{sect:BTZ} we give our Chern-Simons formulation of the
BTZ black hole and in section \ref{sect:MBTZ} we write this
Chern-Simons solution in a more geometric way, and are led to a much
more general solution, which includes the multi-black
hole solutions, some of which have already been discussed by Coussaert
and Henneaux.  We also discuss how gauge transformations act on all
these these solutions.  In section \ref{sect:2BTZ} we specialize to
the case of two black holes (more precisely two excluded regions with
closed timelike curves).  We study the properties of this solution and
the role of degenerate metrics.  We solve the problem of the
non-geodesic singularities of the Coussaert-Henneaux solutions by
choosing a suitable gauge in \ref{subsubsec.Geo} and we end with
conclusions in section \ref{sect:Conc}.

\section{The BTZ black hole}\label{sect:BTZ}
In \(2+1\) dimensional gravity with a negative cosmological constant
there exists a black hole solution to Einstein's equations, the
BTZ black hole, \cite{Banados:1992wn}. It can be viewed either as a metric 
approaching an AdS form asymptotically, or as a
quotient of Anti-de Sitter space \cite{Banados:1993gq}. The BTZ-metric
can be written,
\begin{equation}
\label{ekv:BTZ}
ds^2=-N^2 dt^2+N^{-2 }dr^2+r^2(N^\phi dt+d\phi)^2
\end{equation}
where the lapse function \(N\) and the angular shift \(N^\phi\) are 
\begin{align}
    N^2 &=-M+\frac{r^2}{l^2}+\frac{J^2}{4r^2} & M&=\frac{r_+^2+r_-^2}{l^2}\\
    N^\phi &=-\frac{J}{2r^2} & J&=\frac{2r_+ r_-}{l}\label{MJ}
\end{align}
and \(0<r<\infty\), \(-\infty<t<\infty\), \(0<\phi<2\pi\).  \(M\) is the mass
of the black hole and \(J\) is the angular momentum. Both these
quantities can be expressed in terms of the values of \(r\) (\(r_+\) and
\(r_-\)) when the lapse function \(N\) vanishes. They correspond
to the outer (\(r_+\)) and the inner (\(r_-\)) horizon of the black hole.
For the horizons to exist we need \(M>0\) and \(|J|\leq Ml\). When
\(r_+\) coincides with \(r_-\), we get 
extremal black holes, \(|J|=Ml\). We will be concerned mainly with the 
non-extremal case.

The cosmological constant \(\lambda\) is related to the length scale \(l\)
by the \(\lambda=-1/l^2\).  We choose units such that \(l=1\).
To facilitate the Chern-Simons formulation in section \ref{CSBTZ} we rewrite the 
metric differently the outer
region \(r\!>r\!_+\), the intermediate region \(r_+\!>\!r\!>\!r_-\) and
the inner region \(r_-\!>\!r\!>\!0\). Thus we make the
following Rindler-like coordinate transformation in each region \(r\!>r\!_+\),
\(r_+\!>\!r\!>\!r_-\) and \(r_-\!>\!r\!>\!0\):
\begin{align}
    I: \,\,r^2&=r_+^2 \cosh ^2(\rho-\alpha-\frac{\pi}{2})-r_-^2\sinh
    ^2(\rho-\alpha-\frac{\pi}{2})
    ,\;\;\;\;\; \alpha+\frac{\pi}{2}<\rho<\infty\\
    II: \,\,r^2&=r_-^2 \cos ^2(\rho-\alpha)+r_+^2\sin ^2(\rho-\alpha)
    ,\;\;\;\;\; \alpha<\rho<\alpha+\frac{\pi}{2} \\
    III: \,\,r^2&=r_-^2 \cosh ^2(\rho-\alpha)-r_+^2\sinh ^2(\rho-\alpha)
    ,\;\;\;\;\; 0<\rho<\alpha 
\end{align}
\begin{equation}
\label{alfaBTZ}
\alpha={\mathrm{arctanh}} \left( \frac{r_-}{r_+} \right)
\end{equation}
The constant \(\alpha\) is choosen in such a way that \(r=0\) corresponds
to \(\rho=0\). In these coordinates we get a one to one correspondence
between \(r\) and \(\rho\).
This will lead to the following metrics:
\begin{equation}
  \begin{split}
    I:ds^2=&-\sinh ^2(\rho-\alpha-\frac{\pi}{2})\,
    [r_+dt-r_-d\phi]^2 + d\rho^2 \\
    &+\cosh ^2(\rho-\alpha-\frac{\pi}{2})\,[r_-dt-r_+d\phi]^2\\
    II:ds^2=&\sin ^2(\rho-\alpha)\, [r_-dt-r_+d\phi]^2 - d\rho^2 \\
    &+\cos ^2(\rho-\alpha)\,
    [r_+dt-r_-d\phi]^2\\
    III: ds^2=&-\sinh ^2(\rho-\alpha)\, [r_-dt-r_+d\phi]^2 + d\rho^2 \\
    &+\cosh ^2(\rho-\alpha)\,
    [r_+dt-r_-d\phi]^2   \label{BTZmetric}
  \end{split}
\end{equation}
If we look at the metric in the inner region \(III\) we find that our
choice of \(\alpha\) causes the coefficient of \(d\phi^2\) to vanish
precisely when \(\rho=0\) and to become negative when \(\rho<0\), i.e. we will
get closed timelike curves (CTCs).
Excluding the negative \(\rho\) region in fact removes all CTCs 
\cite{Banados:1993gq}. We also note that \( t \) is always a 
global Killing coordinate, 
timelike in \( I \) and spacelike in \( II \) and \( III \). The \( xy 
\) plane is euclidean in \( I \), 
lorentzian in \( II \) and euclidean in \( III \), implying that light cones are 
drastically tilted inside the black hole. The radial coordinate \( \rho \) 
is spacelike in \( I \), timelike in \( II \) and spacelike in \( III \).

Here it makes sense to pause and think about the split into 
three different coordinate regions. The point we want to make may seem
trivial in the metric formulation, but it will reappear in the 
Chern-Simons formulation. Although the boundaries between 
the regions happen to coincide with the positions of the inner and 
outer horizons there is of course nothing special going on locally 
in these places. So why do we not simply continue our expressions 
from one side of the boundary to the other instead of changing 
analytic forms from region to region? The answer is that the analytic 
expressions of the metric (\ref{BTZmetric}) 
become degenerate at the boundaries of the regions, 
indicating that the coordinates become singular there. In fact, 
if we were to use the region I expression for all \( \rho \) we would 
still have a spacetime divided into two separate regions because 
Einstein's equations of motion cannot really be applied to this
degenerate metric. The true rationale for the matching of different 
metrics across the boundaries between regions is that one can find a 
coordinate chart covering the boundary, and diffeomorphic transformations 
on either side to the respective forms of the metric.

\subsection{Chern-Simons formulation of gravity}
In the Chern-Simons formulation of three-dimensional gravity 
\cite{Achucarro:1986vz} isometries of 
the AdS background are gauged. For AdS the isometry group is \( 
SO(1,2) \times SO(1,2) \) and we call the respective gauge fields of each 
factor \( A = A^k\mathbf{J}_k\) and \( \bar{A} = \bar{A}^k\mathbf{J}_k 
\). The \( SO(1,2) \) generators \( \mathbf{J}_k \) of a factor of the 
group are different from those of the other factor, 
but since they never appear 
multiplied together we shall not distinguish 
between them. The commutation rules within each factor are 
\([\mathbf{J}_k,\mathbf{J}_l]=\epsilon^m{}_{kl}\mathbf{J}_m\), with the
convention \(\epsilon^0{}_{12}=-\epsilon_{012}=-1\), and metric \( 
\eta_{ab} = 2 {\mathrm{Tr}}(\mathbf{J}_a\mathbf{J}_b)\) of signature \( (-1,1,1) \).
The Chern-Simons
three-form 
\begin{equation}
       {\mathrm{Tr}}\left\{A\wedge dA + \frac{2}{3}A\wedge A \wedge A
        \right\}\label{CSForm} 
\end{equation}
and its counterpart for the other factor then serve as Lagrangian 
densities, which automatically yield a 
generally covariant action. The equations of motion 
\begin{equation}
        F= dA + A\wedge A =0 \quad {\mathrm{and}} \quad 
        F= d\bar{A} + \bar{A}\wedge \bar{A}=0
        \label{EqsofM}
\end{equation}
are then actually equivalent with Einstein's equations, provided the 
identifications
\begin{align}
g_{ij}= e_{i}^a e_{j}^b \eta_{ab}\\
 e_{i}^a = \frac{1}{2}\left(A_{i}^a - \bar{A}_{i}^a\right)\\
 \omega_{i}^a = \frac{1}{2}\left(A_{i}^a + \bar{A}_{i}^a\right)
        \label{MetricFromPotential}
\end{align}
of the metric, the dreibein and the spin connection are made, and the
metric is non-degenerate.  Solutions with metrics that are degenerate
somewhere need special study.  In the present paper we encounter cases
where the degeneration corresponds to coordinate singularity or to a
conical singularity.  In some of the cases the degeneration can be
directly associated to horizons, with coordinate singularities in the
accompanying `Schwarzschild-like' coordinate systems.  Such
degenerations may be handled by attaching another coordinate patch
with a boundary and gluing them together by the appropriate matching
conditions.  Then one may find a new coordinate system covering the
boundary region, with a metric which is non-degenerate.  Thus the
degeneration is not a coordinate invariant concept (unless
restrictions are imposed on the allowed coordinate transformations at
a supposed boundary of spacetime).

\subsection{Chern-Simons representation of the BTZ black hole}
\label{CSBTZ}
Now we want to write down the Chern-Simon 
fields corresponding to the
metric in each region, and then verify 
that the field strength \(F\) 
vanishes even at the horizons. We 
need \(F\) to vanish everywhere in the
interior of our space except at singularities for the solutions to 
represent a spacetime with constant negative 
curvature. A non-vanishing field strength at the horizons
can only come from a discontinuity in the \(A\) field when we glue the
different regions together (recall that \(F=dA+A\wedge A\), and if \(A\)
contains a step function the differential gives rise to a
delta function). Since derivatives transverse to the boundary only 
appear in \( F \) for the longitudinal components \( A_{t} \) and \( 
A_{\phi} \), it is enough to ensure that these components are continuous. 

Knowing the metric in the different regions, we may choose
corresponding dreibeins and derive the corresponding spin connections from
the equation of motions, \(de^a+\omega^a{}_b\wedge e^b=0\) and
\(d\omega^a + (1/2)\epsilon^a{}_{bc} \omega^b \wedge
\omega^c=-\frac{1}{2l^2}\epsilon^a{}_{bc} e^b\wedge e^c\). The result 
is unique up to local Lorentz transformations, and a simple choice is
\begin{equation}
\begin{split}
I:\left\{ \begin{array}{l} e=-\sinh
    (\rho-\alpha-\frac{\pi}{2})[r_+dt-r_-d\phi]\mathbf{J}_0+
    \cosh (\rho-\alpha-\frac{\pi}{2})[-r_-dt+r_+d\phi]\mathbf{J}_1 
    + d\rho \mathbf{J}_2 \\
    \omega= - \sinh (\rho-\alpha-\frac{\pi}{2})[-r_-dt+r_+d\phi]\mathbf{J}_0
+ \cosh (\rho-\alpha-\frac{\pi}{2})[r_+dt-r_-d\phi]\mathbf{J}_1
\end{array}
\right. \\ \\
II:\left\{ \begin{array}{l}
 e=d\rho \mathbf{J}_0-\sin
    (\rho-\alpha)[r_-dt-r_+d\phi]\mathbf{J}_1+
 \cos (\rho-\alpha)[r_+dt-r_-d\phi]\mathbf{J}_2 \\
\omega=\sin
    (\rho-\alpha)[r_+dt-r_-d\phi] \mathbf{J}_1 -\cos
 (\rho-\alpha)[r_-dt-r_+d\phi]\mathbf{J}_2
\end{array}
\right. \\ \\
III:\left\{ \begin{array}{l} e=\sinh
    (\rho-\alpha)[-r_-dt+r_+d\phi]\mathbf{J}_0+d\rho \mathbf{J}_1
 +\cosh (\rho-\alpha)[r_+dt-r_-d\phi]\mathbf{J}_2 \\
\omega=\sinh
    (\rho-\alpha)[r_+dt-r_-d\phi]\mathbf{J}_0 +\cosh
 (\rho-\alpha)[-r_-dt+r_+d\phi] \mathbf{J}_2
\end{array}
\right. 
\end{split}
\end{equation}
These dreibeins and spin connections can be compared with those of 
Cangemi et al
\cite{Cangemi:1993my}, who use a different radial coordinate (the same as in
the metric (\ref{ekv:BTZ})). Otherwise the differences are the choices of
some of the signs and in the outer and inner regions 
the Lie algebra components are interchanged. Our choice of 
\( \alpha \) means that closed timelike curves are excluded in 
the region \( 
\rho > 0\), and it corresponds to the boundary condition (at \(\rho 
=0 \)) that the \( \phi 
\)-component of the dreibein is lightlike. In effect it relates the 
tangential components of \( A \) and \( \bar{A} \) at these 
boundaries. 

 From \(A=\omega+e\) and \(\bar{A}=\omega-e\) we get the Chern-Simons 
fields
\begin{equation}
\label{ekv:Afelt}
  \begin{split}
    I:A_\phi=A_{t}&=(r_+-r_-)[\cosh(\rho-\alpha-\frac{\pi}{2})
    \mathbf{J}_1-\sinh(\rho-\alpha-\frac{\pi}{2})\mathbf{J}_0]\\
    II:A_\phi=A_{t}&=(r_+-r_-)[\cos(\rho-\alpha)\mathbf{J}_2+\sin(\rho-\alpha)\mathbf{J}_1]\\
    III: A_\phi=A_{t}&=(r_+-r_-)[\cosh(\rho-\alpha)\mathbf{J}_2+\sinh(\rho-\alpha)\mathbf{J}_0]~,
  \end{split}
\end{equation}
and
\begin{equation}
\label{ekv:barAfelt}
  \begin{split}
    I:\bar{A}_\phi=-\bar{A}_t&=(r_++r_-)[\cosh(\rho-\alpha-\frac{\pi}{2})\mathbf{J}_1
    +\sinh(\rho-\alpha-\frac{\pi}{2})\mathbf{J}_0]\\
    II:\bar{A}_\phi=-\bar{A}_t&=(r_++r_-)[\cos(\rho-\alpha)\mathbf{J}_2-\sin(\rho-\alpha)\mathbf{J}_1]\\
    III:\bar{A}_\phi=-\bar{A}_t&=(r_++r_-)[\cosh(\rho-\alpha)\mathbf{J}_2-\sinh(\rho-\alpha)\mathbf{J}_0]~.
  \end{split}
\end{equation}
Here we see that our choice of dreibeins make the longitudinal 
components of \( A \) and \( \bar{A} \) continuous when
passing between the regions (\(I\rightarrow II\) and so on).
The \(A_\rho\) and the \(\bar{A}_\rho\)
just become
\begin{equation}
\label{ekv:Aro}
  \begin{split}
    I:& A_\rho=\mathbf{J}_2 \;\;\;\; II:A_\rho=\mathbf{J}_0 \;\;\;\; III:A_\rho=\mathbf{J}_1\\
    I:& \bar{A}_\rho=-\mathbf{J}_2 \;\;\;\; II:\bar{A}_\rho=-\mathbf{J}_0 \;\;\;\;
    III:\bar{A}_\rho=-\mathbf{J}_1.
  \end{split}
\end{equation}
Thus the only discontinuous component is \(A_\rho\), which in fact
can not contribute to the field strength since it only depends on 
\( \rho \) and the 
other components are continuous.
\(F_{\rho\rho}\) vanishes by antisymmetry and the off-diagonal terms
\(F_{\phi\rho}\) and \(F_{t\rho}\) vanish by relating the discontinuities 
of \(\partial_{\rho}A_{\phi}\) and  \(\partial_{\rho}A_{t}\) 
respectively with that of \(A_\rho\). 

There is an important distinction between how the boundaries 
between the regions are treated in the Chern-Simons formulation and in 
the metric formulation. In the metric formulation we can be forced to 
match regions with different forms of the metric (or find a 
coordinate patch 
covering the boundary) in order for Einstein's equations to make 
sense everywhere. A naive analytic continuation of
the outer metric (\ref{BTZmetric}) to all \( \rho \) would divide spacetime in two 
disjoint pieces. In contrast, the Chern-Simons formulations seems to leave 
us with a choice. There is nothing wrong with the expressions for the 
vector potentials I, II or III, even if they are extended to all \( 
\rho \). We can take those expressions as they are (giving us a problem in 
the gravitational interpretation) or we can match solutions and get 
the BTZ solution.

From a Chern-Simons perspective the matched discontinuous solutions 
and the smooth solutions are indistuinguishable in the outer region, 
and they both make
equally good sense in the interior. Only imposing boundary conditions 
in the interior or imposing special gauge conditions may pick out one 
solution as preferable to the other. Thus a sound gravitational 
interpretation of the solutions is only possible given special 
boundary conditions or gauge fixings of the vector potential. In 
generalizing the BTZ solution we will ensure that the boundaries of 
different regions are always matched in the same way as in this original 
BTZ solution.

To prepare for more general solutions let us write the BTZ solution in
cartesian coordinates, \(\rho=\sqrt{x^2+y^2}\) and
\(\phi=\arctan\left(\frac{y}{x}\right)\).  In the inner region we can
write it as,
\begin{align}
    A_x&=\mathbf{g}\frac{-qy}{x^2+y^2}+\partial_x \rho \; \mathbf{J}_1\\
    A_y&=\mathbf{g}\frac{qx}{x^2+y^2}+\partial_y \rho \; \mathbf{J}_1
    \label{btz-solution}
\end{align}
where
\begin{equation}
q=r_+-r_-\label{q}
\end{equation}
and
\begin{equation}
    \mathbf{g}=[\cosh(\rho-\alpha)\mathbf{J}_2+\sinh(\rho-\alpha)\mathbf{J}_0].
\end{equation}
The second vector potential
  \begin{align}
    \bar{A}_x&=\mathbf{\bar{g}}\frac{-\bar{q}y}{x^2+y^2}-\partial_x \rho \; \mathbf{J}_1\\
    \bar{A}_y&=\mathbf{\bar{g}}\frac{\bar{q}x}{x^2+y^2}-\partial_y \rho \; \mathbf{J}_1
  \end{align}
where
\begin{equation}
\bar{q}=r_++r_-\label{qbar}
\end{equation}
and
\begin{equation}
    \mathbf{\bar{g}}=[\cosh(\rho-\alpha)\mathbf{J}_2-\sinh(\rho-\alpha)\mathbf{J}_0].
\end{equation}
In cartesian coordinates it looks as if \( \rho = 0 \) denotes a 
single point in space. There is no a priori justification for this since 
we chose \(\rho = 0\) to be special by hand, and all other equations \( 
\rho = {const \rm} \) denote topological circles. On the other hand, we excluded \(\rho\leq 0\)
on physical grounds, to get rid of closed timelike curves.  
Furthermore, calculating \( F \) in cartesian coordinates we get a 
delta function at the origin which we may formally regard as a 
source, and in this context we can also regard \( \rho = 0 \) as a 
single point.

\subsection{Holonomies}
In a gauge theory of flat connections, \( F= \bar{F}= 0 \), gauge 
invariant observables are scarce. The fields are locally pure gauge 
\(A = U^{-1}dU\), for \( U \) an element of \(SO(1,2)\),
and any non-trivial observable has to be associated with the boundaries 
of spacetime or be topological in nature. The simplest topological 
observables are holonomies (or Wilson loops) measuring the effect of 
parallel transport along a closed loop in spacetime. For flat 
connections the result can only be non-zero if the loop \( C_{x} \) 
(based at \( x \)) is 
non-contractible. Then the Wilson loop
\begin{equation}
     W(C_{x})={\mathcal{P}}\exp\left(\oint_{C_{x}}A\right)=U^{-1}(x)U(x+C_{x})~,
\end{equation}
where \( {\mathcal{P}} \) denotes path ordering of the exponential.
As observed by Cangemi et al. \cite{Cangemi:1993my} 
it is simplest in our case to 
take the closed 
curve \( C_{x} \) at 
constant radial coordinate, i.e. along a level curve of \( \rho \). 
For two curves \( C_{x}\) and \(C_{y}\) which can be continuously deformed 
into each other, but are based at two different points \(x\) and \(y\), 
the holonomies are conjugate, \(W(C_{y}) = U(y)^{-1} U(x) W(C_{x}) U(x)^{-1} 
U(y)\). The \emph{eigenvalues} of \( W \) for two 
curves which can be continuously deformed into each other are thus 
equal. These eigenvalues 
are determined by the parameters \( q \), 
\( \bar{q} \) and the eigenvalues of the 
\(SO(1,2)\) Lie algebra elements \( \mathbf{g}\) and \( \mathbf{\bar{g}} \). It does 
not matter 
in which coordinate patch we follow the level curves, because we have 
ensured that 
the connections are flat also at the boundaries between the patches. 

Since the gauge group is a product of two rank one groups it is 
enough to characterize the eigenvalues by the two traces 
\({\mathrm{Tr }}\:W(C) \) and \({\mathrm{Tr}}\:\bar{W}(C) \). For the BTZ 
solutions we obtain 
\begin{equation}
    {\mathrm{Tr}}\:W(C) = 2 \cosh[\pi q],\qquad 
    {\mathrm{Tr}}\:\bar{W}(C) = 2 \cosh[\pi \bar{q}],
    \label{eq:Hol}
\end{equation}
for Wilson loops in the two-dimensional representation of \(SO(1,2)\). 
Via Equation \ref{MJ} the holonomies are then related to the 
mass and spin of the black hole. In the complete classification 
of conjugacy classes of \(SO(2,2)\) Lie algebra elements 
\cite{Banados:1993gq} one finds 
that holonomies corresponding formally to imaginary \(q\) or \( 
\bar{q}\) may occur, and furthermore that the case of coinciding 
eigenvalues (when \({\mathrm{Tr}}\:W(C) = 2 \) or 
\({\mathrm{Tr}}\:\bar{W}(C) = 2\)) allows for non-trivial 
holonomy matrices (in addition to \( W = 1\) or \(\bar{W}=1 \)). These 
cases can be dealt with in the 
Chern-Simons formulation by modifying the expressions for \( \mathbf{g} \) and 
\( \mathbf{\bar{g}} \).

\section{Multi-black hole solutions}\label{sect:MBTZ}
We will generalize the solution (\ref{btz-solution}) to the case were
we have arbitrary many singularites. We will use the same form of
the solution as in the inner region, regarding the Lie algebra
direction of A.\footnote{If we exchange \(\mathbf{J}_1\) with \(\mathbf{J}_2\) we also
have to change signs in front of the \(\mathbf{J}_0\) component, in 
order to preserve the commutation relations which govern the equations 
of motion. If we exchange
\(\mathbf{J}_1\) with \(\mathbf{J}_0\) we have to change the last condition in
(\ref{eq:hyperbolic}) below. There will then be a minus sign in front 
of \(g_1\), in effect exchanging trigonometric and hyperbolic functions. This
is precisely the case in (\ref{ekv:Afelt}), (\ref{ekv:barAfelt}) and
(\ref{ekv:Aro}). } We may then try a solution 
\begin{align}
    A&=dh \mathbf{J}_{1}+ (f+dt) 
    \mathbf{g}\label{ansatz1}\\
    \mathbf{g}&=g_{0}(h) \mathbf{J}_{0} + g_{2}(h) \mathbf{J}_{2}~,
    \label{ansatz2}
\end{align}
where \( h \) is a scalar function generalizing the radial coordinate 
\( \rho \) and \( f \) is a spatial one-form which is closed except at 
isolated sources
\begin{equation}
    df = 2 \pi\sum_{i=1}^{N} q_i \delta^2(\vec{x}-\vec{x}_i)\; dx\wedge 
    dy~.\label{eq:f}
\end{equation}
The \(q_{i}\) determine the strength of the sources (the masses 
and spins of black holes). By integrating (\ref{eq:f}) over a large 
disk \( D \) enclosing all sources we obtain
\begin {equation}
\oint_{\partial D} f =\int_{D} df = 2 \pi \sum_{i=1}^{N} q_i  = 2\pi Q~.
\end{equation}
If appropriate boundary conditions on \( f \) are assumed, \( f \to Q 
d\phi \) as \( \rho \to \infty \). Then we may regard \(A_{t}= Q 
\mathbf{g}(h) \to A_{\phi} \) as a natural generalization of the 
relation \( 
A_{t}-A_{\phi}=0\) satisfied by single BTZ black holes. This is 
consistent with the ansatz (\ref{ansatz1}) after rescaling \(t\).

The equations of motion \( dA + A \wedge A=0 
\) are satisfied by the vector potential (\ref{ansatz1}) outside the sources, 
\( \vec{x}\neq 
\vec{x}_i)\), 
provided
\begin{equation}
    \frac{dg_{0}}{dh}=g_{2}~,\qquad \frac{dg_{2}}{dh}=g_{0}~.
    \label{eq:hyperbolic}
\end{equation}
We recognize the equation for the hyperbolic functions entering the 
BTZ solution, but now their arguments have been generalized from \( \rho \) 
to \( h \). By permuting the Lie algebra elements \( \mathbf{J}_{i} \) in 
Equations (\ref{ansatz1}, \ref{ansatz2}) one obtains solutions generalizing 
the BTZ solutions 
for all three regions, provided the signs in Equations 
(\ref{ansatz1}, \ref{ansatz2}, \ref{eq:hyperbolic}) are changed 
accordingly. Matching of the regions works precisely as in the BTZ case. Note 
that the Lie algebra element \( 
\mathbf{g} \) is spacelike, null or timelike depending on the sign of 
\( {\mathrm{Tr}}\:\mathbf{g}^2\), and that its sign is necessarily constant all over 
spacetime for the present solutions.  
The one-form \( f /Q \) generalises the angular one-form \( d\phi \) in 
the BTZ case. 

Although any choices of \( h \) and of \( f \) satisfying 
Equation (\ref{eq:f}) are consistent with the equations of motion, we 
will concentrate on boundary conditions and combinations of \(A\) and 
\(\bar{A}\) solutions that reduce to ordinary BTZ solutions both for 
asymptotically large \( h \) and close to the sources (regions of 
closed timelike curves). All the important new features of these generalized 
solutions are then associated with the fact that they are 
multi-centered, which in its turn implies that there will be critical points of \( 
h \) and \( f \). Such critical points can give rise to degenerate 
metrics, a subject we shall return to in Section (\ref{subsec.DegMet}).

The second gauge field \(\bar{A}\) has analogous solutions in terms 
of \(\bar{h}\),
\(\bar{g}_2(\bar{h})\) and \(\bar{g}_0(\bar{h})\).
In order to get solutions similar to the BTZ solutions we may choose
\(\bar{h}=-h\),
\(\bar{g}_2(\bar{h})=g_2(\bar{h})\) and 
\(\bar{g}_0(\bar{h})=g_0(\bar{h})\), guided by 
Equations (\ref{ekv:Afelt}) and (\ref{ekv:barAfelt}).
In an inner region with \(g_{2}\) and \(g_{0}\) even and odd 
functions respectively, we obtain the vector potentials
\begin{equation}
\begin{split}
A&=(f+Qdt)g_0(h)\mathbf{J}_0+dh \mathbf{J}_1+
(f+Qdt)g_2(h)\mathbf{J}_2\\
\bar{A}&=-(\bar{f}-\bar{Q}dt)g_0(h)\mathbf{J}_0-dh \mathbf{J}_1+
(\bar{f}-\bar{Q}dt)g_2(h)\mathbf{J}_2
\end{split}
\end{equation}
and the metric,
\begin{equation}
ds^2=-g_0^2(h(x,y))\left\{r_- dt - f_{+}(x,y)\right\}^2
    +g_2^2(h(x,y))\left\{ r_+ dt -  f_{-}(x,y) \right\}^2
	+dh(x,y)^2~,\label{SimpMultiBTZMet}  
\end{equation}
where
\begin{equation}
    f_{\pm}=\frac{\bar{f}\pm f}{2}~,\qquad r_{\pm}=\frac{\bar{Q}\pm Q}{2}
    \label{eq:f+-}
\end{equation}
The metric is easily compared with the BTZ metric 
(\ref{BTZmetric}) in the inner region (\(III\)). 
The function \(h + \alpha\) 
has replaced the radial coordinate \( \rho \), \( g_{0} \) and \( 
g_{2} \) represent the hyperbolic functions, and \(r_{\pm} d\phi \) is 
replaced by \( f_{\pm} \). The last change is the most significant one,
since two different one-forms are needed to generalize \( d\phi \). 
Only when \( f_{+} \) and  \( f_{-} \) are proportional do we get a 
direct multi-source generalization of \( d\phi \). This happens when the 
ratio of the two charges at each source is constant.
Irrespective of this we can make direct contact 
with the BTZ-solution very close to a charge, where the effect of the other charges is 
negligable, or at asymptotically large distances, where the 
sum of the charges dominate the solution. 

In the general case (\ref{SimpMultiBTZMet}) we can still define
regions of type \(I\), \(II\) and \(III\), between which the solutions
have to be matched, and different choices of the function \( h\) gives
different regions (even their topologies may be different), but they
are actually related by gauge transformations, as we proceed to 
discuss.

\subsection{Gauge transformations} \label{subsec.Gauge}

One can check that the gauge transformation
\begin{equation}
    \delta A =d \delta h  \mathbf{J}_{1}
    + \left[A, \delta h  \mathbf{J}_{1}  \right]
    \label{eq:h-shift}
\end{equation}
amounts to a change of \( h \) into \( h + \delta h \) in the solution 
for region \(III\), implying that 
solutions with different functions \( h \) are equivalent if only 
their boundary conditions are the same. Of course, an analogous 
statement is true for \(\bar{A}\). We stress that the solutions are only 
equivalent in the Chern-Simons formulation of pure gravity. To see 
this we may study horizons.

The boundary between region \(I\) and \(II\) resembles a horizon 
and it is actually an event horizon for constant charge ratios of all sources, 
if it consists of a single connected component. This is because we can 
then use coordinates \(h\) and \(\psi\) with \(d\psi = f\) to obtain the 
ordinary BTZ metric in the exterior region. The transformation to 
these coordinates works asymptotically and also in the whole exterior 
region provided \(f\) does not have 
a zero there. In fact, we will show 
later that the multi-black hole solutions have 
singularities at zeroes of \(f\). These singularities may be inside or outside 
a physical event horizon 
depending on the choice of the function \( h \). Such a difference 
could for instance be detected by the propagation of light rays in the 
background metric. Light rays are of course not included in a 
Chern-Simons description.

Even if there is little physics in the function \( h \), the 
multi-black hole solution also depends on the forms \( f_{+} \) and 
\( f_{-} \), which in their turn depend on the positions and charges 
of the sources. As will be discussed in the next subsection, the 
charges may be directly measured by holonomies around the sources. The 
positions of the sources are trickier, and cannot be resolved by the 
holonomies. Other available 
observables are the asymptotic charges \cite{Banados:1996tn}. 
The general \( f_{+} \) and \( f_{-} \) 
are asymptotic to the corresponding BTZ forms, and the issue is if the 
approach is fast enough to give finite asymptotic charges, but also 
slow enough to give non-zero values. 

As an example we may compare a single source BTZ 
solution \( A_{1} \) with a solution \( A_{2} \) with sources
separated by a small 
coordinate distance \( x_{0} \) in the \( x \) direction. Then one 
finds
\begin{equation}
    A_{2} =  A_{1} + \delta_{12} A_{1} =  A_{1} +
    x_{0} (q_{2}-q_{1}) d \left(\frac{-y}{x^2+y^2}\right)\mathbf{g}~.
    \label{eq:A-shift}
\end{equation}
Thus \( \delta_{12} A_{1\phi} \) scales as \( \mathbf{g}\rho^{-2} \) 
while \( A_{1\phi} \) scales as \( \mathbf{g}\rho^{-1} \) with \( 
\rho \) and the change is subleading. If the change \( \delta_{12} 
A_{1\phi} \) can be written as an infinitesimal gauge transform \( 
\delta_{\Lambda_{12}} A_{1} \) with a decreasing gauge parameter \( 
\Lambda \) then the separation of the two sources is truly a matter of 
gauge choice at infinity and it is not detectable by any asymptotic 
charges. (It will still be detectable by holonomies, corresponding to 
the fact that the gauge transformations are not defined everywhere, or 
do not belong to the identity component of the gauge group.) The 
problem in our case is that the asymptotic behaviour of the BTZ 
solution implies that \( \mathbf{g} \) has an exponential dependence 
on \( \rho \). The same is true for \( \Lambda_{12} \). Then the 
boundary values of the fields and the transformation parameters are 
not well defined. Fortunately, this problem may be circumvented by 
discussing the vector potentials
\begin{equation}
     A' = e^{\rho \mathbf{J}_{2}} d e^{ - \rho \mathbf{J}_{2}} 
     + e^{\rho \mathbf{J}_{2}} A e^{ -\rho \mathbf{J}_{2}}
     \label{A'}~,
\end{equation}
which locally are gauge transforms of \( A \) but satisfy different
boundary conditions. In fact \( A'_{BTZ} \) is a constant and the \( 
A' \) of our 
generalized multi-source solutions approach constants at infinity. The 
\( A' \) do however give rise to metrics which are everywhere degenerate, and we 
just regard them as auxiliary solutions which help distinguishing asymptotic gauge 
transformations and global transformations generated by asymptotic 
charges. The parameters of global transformations on \( A' \) go to 
constants at infinity while true gauge transformations vanish 
asymptotically. The effect of both kinds of transformations on the 
fields \( A \) is simply obtained by the mapping inverse to 
(\ref{A'}). Conversely, by mapping to \( A' \) transformations on \( 
A \) may be classified as gauge transformations or global 
transformations (or as changing boundary conditions).

Returning to \( \delta_{12} A_{1\phi} \), its image \( \delta_{12} 
A'_{1\phi} \) under the map 
(\ref{A'}) vanishes at infinity, implying 
that asymptotic charges are left invariant by moving sources apart. In fact, 
\( \delta_{12} A'_{1\phi} = \delta_{\Lambda_{12}} A_{1}\) for \( \Lambda_{12} = 
-y\mathbf{g'}/(x^2+y^2)  \) with a constant \(\mathbf{g'}\).  Since 
\(\Lambda_{12} \) diverges at the origin it does not give a globally 
well defined infinitesimal gauge transformation and there can still 
be a physical difference between the solutions. In 
conclusion, solutions with different numbers of sources are 
inequivalent because of different holonomies, while different positions 
of the sources may or may not
be observable depending on the global properties and boundary 
conditions of the finite gauge 
transformations effecting the translations. The asymptotic charges 
are insensitive to these details, so they may be thought of as generating 
transformations common to several different sectors labeled by the numbers of sources, and 
possibly by their positions.

\subsection{Multi-black hole holonomies}
We now wish to calculate holonomies 
\begin{equation}
{\mathrm{Tr}}\:\left( \mathcal{P}\exp{\int_C A}\right)
\end{equation}
We first calculate the ordinary integral over a closed loop,
\begin{equation}
     \int_C A=\int
    \left(\left(f_x \mathbf{g}+\frac{\partial h(x,y)}{\partial 
    x}\mathbf{J}_{1}\right) dx +
      \left(f_y \mathbf{g}+\frac{\partial h(x,y)}{\partial y}\mathbf{J}_{1}\right) dy 
    \right) ~,
 \end{equation}
 here written out for an inner-type region. 
 If the function \(h(x,y)\) is choosen in such a way that there are
 closed level curves of \(h(x,y)\) the term \(dh\) in the integral is
 zero, and furthermore \( \mathbf{g} \) is constant.  
Then the integral depends on which charges \(q_{i}\) are enclosed by the level 
curve,
\begin{equation}
  \int_C A=\mathbf{g} \int_C f = 2 \pi \mathbf{g} \sum_{i \in I_{C}} 
  q_{i}= 2 \pi \mathbf{g} q_{C}~,
\end{equation}
where \( I_{C}\) denotes the set of 
enclosed sources and \( q_{C} \) the enclosed charge. 
Since the eigenvalues of the traceless real matrix \(\mathbf{g}\)
are necessarily both real or both imaginary and add up to zero, we 
may write
\begin{equation}
  {\mathrm{Tr}}\:\left( \mathcal{P}\exp{\int_C A}\right) =
  e^{2\pi q_C\lambda}+e^{-2\pi q_C\lambda}= 2 \cosh{(2\pi q_C\lambda)}
\end{equation}
where \(\lambda\) is one of the eigenvalues of 
\(\mathbf{g}\), and independent of \(h\). 
The matrix corresponding to the \(\bar{A}\) must also have either both
imaginary or both real eigenvalues which we call \(\bar{\lambda}\) and
\(-\bar{\lambda}\). So in general we get three different holonomy types 
depending on the eigenvalues \(\lambda\) and \(\bar{\lambda}\):
either one is real and one imaginary, both are real or both are 
imaginary. 
When we just have one singularity it is known that these types will
correspond to different quotients of anti-de Sitter space. Ba\~{n}ados et 
al
\cite{Banados:1993gq} have shown how different spaces are obtained
from anti-de Sitter by modding out subgroups of \(SO(2,2)\), and that
BTZ black holes belong to one of these classes of spaces. They also 
find three different types of spaces. The correspondence between their
eigenvalues \(\lambda'\) and our eigenvalues is
\(\lambda_1'=q\lambda-\bar{q}\bar{\lambda}\) and
\(\lambda_2'=q\lambda+\bar{q}\bar{\lambda}\). In our language the 
generic BTZ black hole corresponds to the case with two real 
eigenvalues. When both are imaginary we generally get conical 
singularities, except in the case \( q \lambda=\bar{q}\bar{\lambda}=i/2\), 
which curiously corresponds to AdS space\footnote{\(M=-1\) and \(J=0\) are 
obtained from Equations (\ref{q}, \ref{qbar}) and (\ref{MJ}) and the metric 
(\ref{ekv:BTZ}) 
then represents AdS.}. In fact, we may also find `multi-AdS solutions' 
with several of these AdS charges. They may possibly serve as ground 
states of multi-black hole sectors. Note that the holonomy around a 
single AdS charge is almost trivial, and around two it is entirely 
trivial.

Notice that we have not mixed holonomy type for the different
singularities. It would be interesting 
to find solutions where the sources give rise to different types of
holonomies.

\section{Two sources}\label{sect:2BTZ}
We will study the solutions  for the case with two
sources in more detail. After verifying that the solutions approach 
the single-source solution asymptotically and for vanishing 
separation of the charges, we will continue with a generalization to 
several sources of the 
procedure to exclude CTCs, and we will also discuss how the 
multi-source solutions generically contain additional (mild) 
singularities.

Solutions with sources at \( x=x_{1}=x_{0}\) and \(x=x_{2}=-x_{0}\) can be written,
  \begin{align}
  A_t&=[(r_{1+}-r_{1-})+(r_{2+}-r_{2-})] \mathbf{g}\\
   A_x&=(f_{1x}+f_{2x})\mathbf{g}+\frac{\partial h(x,y)}{\partial x} \mathbf{J}_1\\
    A_y&=(f_{1y}+f_{2y})\mathbf{g}+\frac{\partial h(x,y)}{\partial y} \mathbf{J}_1
  \end{align}
where
\begin{align}
  f_{1x}&=q_{1}\frac{-y}{(x-x_0)^2+y^2}  \;\; 
  f_{1y}=q_{1}\frac{x-x_0}{(x-x_0)^2+y^2}\\
  f_{2x}&=q_{2}\frac{-y}{(x+x_0)^2+y^2}  \;\; 
  f_{2y}=q_{2}\frac{x+x_0}{(x+x_0)^2+y^2}~.
\end{align}
The conjugate field \(\bar{A}\),
  \begin{align}
    \bar{A}_x&=(\bar{f}_{1x}+\bar{f}_{2x})\bar{\mathbf{g}}+
    \frac{\partial \bar{h}(x,y)}{\partial x} \mathbf{J}_1\\
    \bar{A}_y&=(\bar{f}_{1y}+\bar{f}_{2y})\bar{\mathbf{g}}+
    \frac{\partial \bar{h}(x,y)}{\partial y} \mathbf{J}_1
  \end{align}
where,
  \begin{equation}
    \bar{\mathbf{g}}=\bar{g}_2(\bar{h}(x,y))\mathbf{J}_2+
    \bar{g}_0(\bar{h}(x,y))\mathbf{J}_0~.
  \end{equation}
In the BTZ-like inner region with \(\bar{h}=-h\), 
\(\bar{g}_2(\bar{h})=g_2(\bar{h})\), 
\(\bar{g}_0(\bar{h})=g_0(\bar{h})\) and 
\(g_{2}\) and \(g_{0}\) even and odd 
functions respectively, 
we find the metric
\begin{equation}
  \begin{split}
    ds^2=&-g_0^2(h(x,y))\left\{(r_{1-}+r_{2-})dt-(r_{1+}f_{1x}+r_{2+}f_{2x})dx
      -(r_{1+}f_{1y}+r_{2+}f_{2y})dy \right\}^2\\
    &+g_2^2(h(x,y))\left\{ (r_{1+}+r_{2+})dt-(r_{1-}f_{1x}+r_{2-}f_{2x})dx
      -(r_{1-}f_{1y}+r_{2-}f_{2y})dy \right\}^2\\
  &+\left\{\frac{\partial h(x,y)}{\partial x}dx+ \frac{\partial
        h(x,y)}{\partial x}dy \right\}^2~.
  \end{split}
\end{equation} 

So far the function \( h \) has been left unspecified. If, for 
instance, we choose  
\(h(x,y)=\sqrt{\rho_1 \rho_2}-\alpha\) in terms of the radial 
coordinates \( \rho_1\) and \( \rho_2\) centered on each of the two sources and a 
function \(\alpha\) approaching a constant (\ref{alfaBTZ}) at infinity 
and in the limit \(x_0\rightarrow 0\), we can ensure that the BTZ solution 
is approached both at infinity and as \(x_0\rightarrow 0\). To verify 
this, start by looking at the metric in the outer region 
\begin{equation}
 \begin{split}
    ds^2=&-\sinh^2(\sqrt{\rho_1
    \rho_2}-\alpha)\left((r_{1+}+r_{2+})dt-\left((r_{1-}f_{1x}+r_{2-}f_{2x})dx
    +(r_{1-}f_{1y}+r_{2-}f_{2y})dy\right)\right)^2\\ 
    &+\cosh^2(\sqrt{\rho_1
    \rho_2}-\alpha)\left((r_{1-}+r_{2-})dt-\left((r_{1+}f_{1x}+r_{2+}f_{2x})dx
    +(r_{1+}f_{1y}+r_{2+}f_{2y})dy\right)\right)^2\\
    &+\left(\frac{\partial h(x,y)}{\partial x}dx+ \frac{\partial
    h(x,y)}{\partial x}dy \right)^2~,
 \end{split}
\end{equation}
to see how it
behaves asymptotically at infinity.
In terms of polar coordinates \( (\rho,\phi)\) centred around \((x,y) = 
(x_1,0)\), (implying \( \rho = \rho_{1} \) )
\begin{equation}
  \begin{split}
    f_{1x}&=\frac{-y}{\rho^2}=\frac{-\rho\sin \phi}{\rho^2} \ \ \ \
    \ \ \ \ \ \ 
    f_{1y}=\frac{x-x_0}{\rho^2}=\frac{\rho\cos \phi}{\rho^2}\\
    f_{2x}& =\frac{-y}{\rho_2^2}=\frac{-\rho\sin \phi}{\rho_2^2} \ \ \
    \ \ \ \ \ \ \ 
    f_{2y}=\frac{x+x_0}{\rho_2^2}=\frac{\rho\cos \phi +2x_0}{\rho_2^2}
  \end{split}
\end{equation}
the metric takes the form
\begin{equation}
  \begin{split}
    ds^2_{\mbox{outer}}=&-\sinh^2(\sqrt{\rho
      \rho_2}-\alpha)\left\{r_{+}dt- 
    \left(\frac{r_{2-}(\rho^2+2x_0\rho\cos
        \phi)+r_{1-}\rho_2^2}{\rho_2^2} d\phi  
    +\frac{2r_{2-}x_0\sin \phi}{\rho_2^2}d\rho\right)\right\}^2\\ 
    &+\cosh^2(\sqrt{\rho \rho_2}-\alpha)\left\{r_{-}dt-
    \left(\frac{r_{2+}(\rho^2+2x_0\rho\cos \phi)+r_{1+}\rho_2^2}{\rho_2^2} d\phi
    +\frac{2r_{2+}x_0\sin \phi}{\rho_2^2}d\rho\right)\right\}^2\\ 
    &+\left\{\frac{\partial h(\phi,\rho)}{\partial \phi}d\phi+ \frac{\partial
        h(\phi,\rho)}{\partial \rho}d\rho \right\}^2.\label{Outer2BH}
  \end{split}
\end{equation}
We see that the metric is asymptotic to the BTZ solution with \(r_+=r_{1+}+r_{2+}\) and
\(r_-=r_{1-}+r_{2-}\) when
\(\rho\rightarrow \infty\) or \( x_{0}\rightarrow 0 \).

\subsection{Exclusion of closed timelike curves}

In the BTZ solution (\ref{BTZmetric}) there are closed timelike curves for 
\( \rho < 0 \), and we expect similar pathologies in the multi-black 
hole solutions inside the black holes. It is natural to cut off the 
range of the coordinates precisely where CTCs are encountered. Here 
we show how this can be done in the case of two sources. The same procedure 
can be used for any number of sources. The resulting
spacetimes then have singularities in the causal structure if 
they are continued `inside' the sources. 

Just as 
for the BTZ case (\ref{BTZmetric}) we need
the vector field \(\partial_\phi\) for some periodic
coordinate \(\phi\) to become lightlike at each
source in order to exclude regions containing closed timelike
curves. Coordinates which are periodic around curves enclosing only single 
sources are readily found. We may use the angle between the line from 
the source to a point and the 
positive 
\(x\) direction, or we may use \( df_{+}\) and \(df_{-}\) to measure 
angular differences. Close to the sources these measures of angle
all agree up to proportionality 
constants. 

To localize the causal singularities to the positions of the sources 
it is then enough to choose the function \( \alpha \) appropriately. 
In order to encounter closed timelike curves we have to go to the inner 
region. 

First study the metric in the inner region. It
is obtained from the outer metric (\ref{Outer2BH}) by exchanging \(r_+\) with \(r_-\):
\begin{equation}
  \begin{split}
    ds^2_{\mbox{inner}}=&-\sinh^2(\sqrt{\rho
      \rho_2}-\alpha)\left\{r_{-}dt- 
    \left(\frac{r_{2+}(\rho^2+2x_0\rho\cos
        \phi)+r_{1+}\rho_2^2}{\rho_2^2} d\phi  
    +\frac{2r_{2+}x_0\sin \phi}{\rho_2^2}d\rho\right)\right\}^2\\ 
    &+\cosh^2(\sqrt{\rho \rho_2}-\alpha)\left\{r_{+}dt-
    \left(\frac{r_{2-}(\rho^2+2x_0\rho\cos \phi)+r_{1-}\rho_2^2}{\rho_2^2} d\phi 
    +\frac{2r_{2-}x_0\sin \phi}{\rho_2^2}d\rho\right)\right\}^2\\ 
    &+\left\{\frac{\partial h(\phi,\rho)}{\partial \phi}d\phi+ \frac{\partial
        h(\phi,\rho)}{\partial \rho}d\rho \right\}^2.
  \end{split}
\end{equation}
Now take a look at the \(g_{\phi \phi}\) component,
\begin{equation}
  \begin{split}
    g_{\phi \phi}=&- \left(\frac{r_{2+}(\rho^2+2x_0\rho\cos \phi)+r_{1+}
        \rho_2^2}{\rho_2^2} \right)^2
    \sinh^2(\sqrt{\rho \rho_2}-\alpha) \\
    &+ \left(\frac{r_{2-} (\rho^2+2x_0\rho\cos
        \phi)+r_{1-}\rho_2^2}{\rho_2^2} \right)^2
    \cosh^2(\sqrt{\rho \rho_2}-\alpha)\\
    &+\frac{x_0^4\rho\sin^2\!2\phi}{\rho_2^3}+ \left(\frac{\partial
        \alpha}{\partial \phi}\right)^2+
    2\frac{x_0^2\sqrt{\rho}\sin 2\phi}{\rho_2^{3/2}}
    \left(\frac{\partial\alpha}{\partial \phi}\right),
  \end{split}
\end{equation}
when \(\rho=0\). We must choose
\(\alpha\) in order to make the vector field
\(\partial_\phi\) lightlike at \(x=x_0\). For \(\alpha\) with
\(\partial_\phi \;\alpha=0\) when \(\rho=0\), the condition that \(\partial_\phi\)
becomes lightlike becomes
\begin{equation}
\begin{split}
g_{\phi \phi}&=-(r_{1+})^2\sinh^2\alpha+(r_{1-})^2\cosh^2\alpha=0 \\
\Rightarrow \alpha&={\mathrm{arctanh}} \left( \frac{r_{1-}}{r_{1+}} \right).
\end{split}
\end{equation}
In the same way we can change to polar coordinates centred around 
\(x=-x_{0}\)
which instead would lead us to the condition,
\begin{equation}
\begin{split}
g_{\phi \phi}&=-(r_{2+})^2\sinh^2\alpha+(r_{2-})^2\cosh^2\alpha=0 \\
\Rightarrow \alpha&={\mathrm{arctanh}}\left( \frac{r_{2-}}{r_{2+}} \right).
\end{split}
\end{equation}
In order to have both these conditions satisfied \(\alpha\) can only be
a constant in the case \(r_{1-}/r_{1+}=r_{2-}/r_{2+}\). Still, there are 
many
ways of choosing an \(\alpha (\rho,\phi)\) that does not affect the
singularities or the asymptotics of the solutions. We may choose
\(\alpha\) to be a constant at infinity, for instance 
\begin{equation}
  \alpha={\mathrm{arctanh}} \left(\frac{r_{1-}\rho_2+r_{2-}\rho_1}
{r_{1+}\rho_2+r_{2+}\rho_1} \right)
\end{equation}
We see that in the case \(r_{1-}/r_{1+}=r_{2-}/r_{2+}\) this \(\alpha\) will 
reduce to a constant. This will also be
the case when \(x_0=0\), i.e. when the singularities are in the same
point. The requirement \(\partial_\phi\,\alpha=0\) when \(\rho=0\) is also
easily seen to be fulfilled.

\begin{figure}
    \rotatebox{-90}{
    \epsfig{ figure = 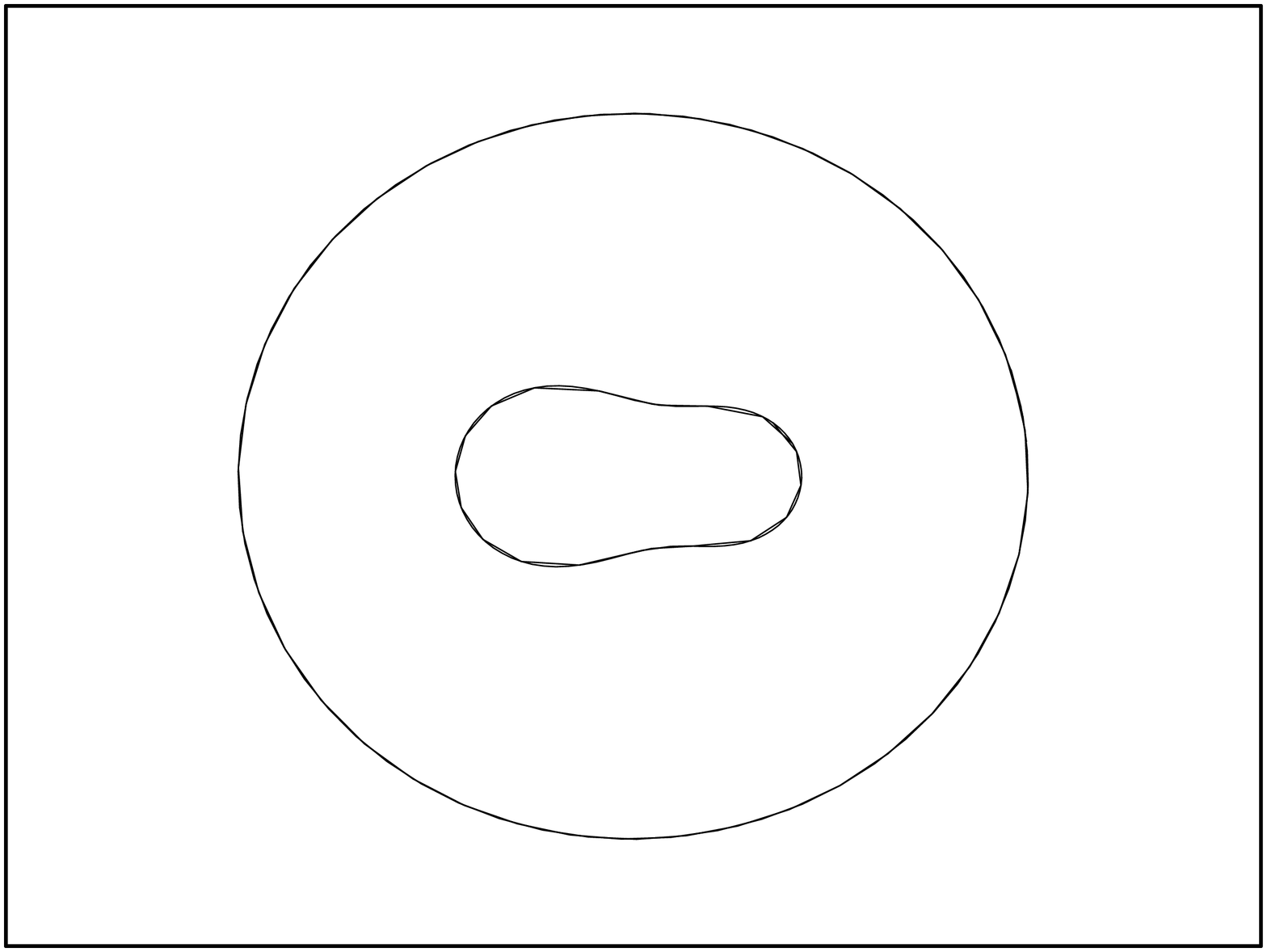, width=5cm, height=4.8cm }}
    \rotatebox{-90}{
    \epsfig{ figure = 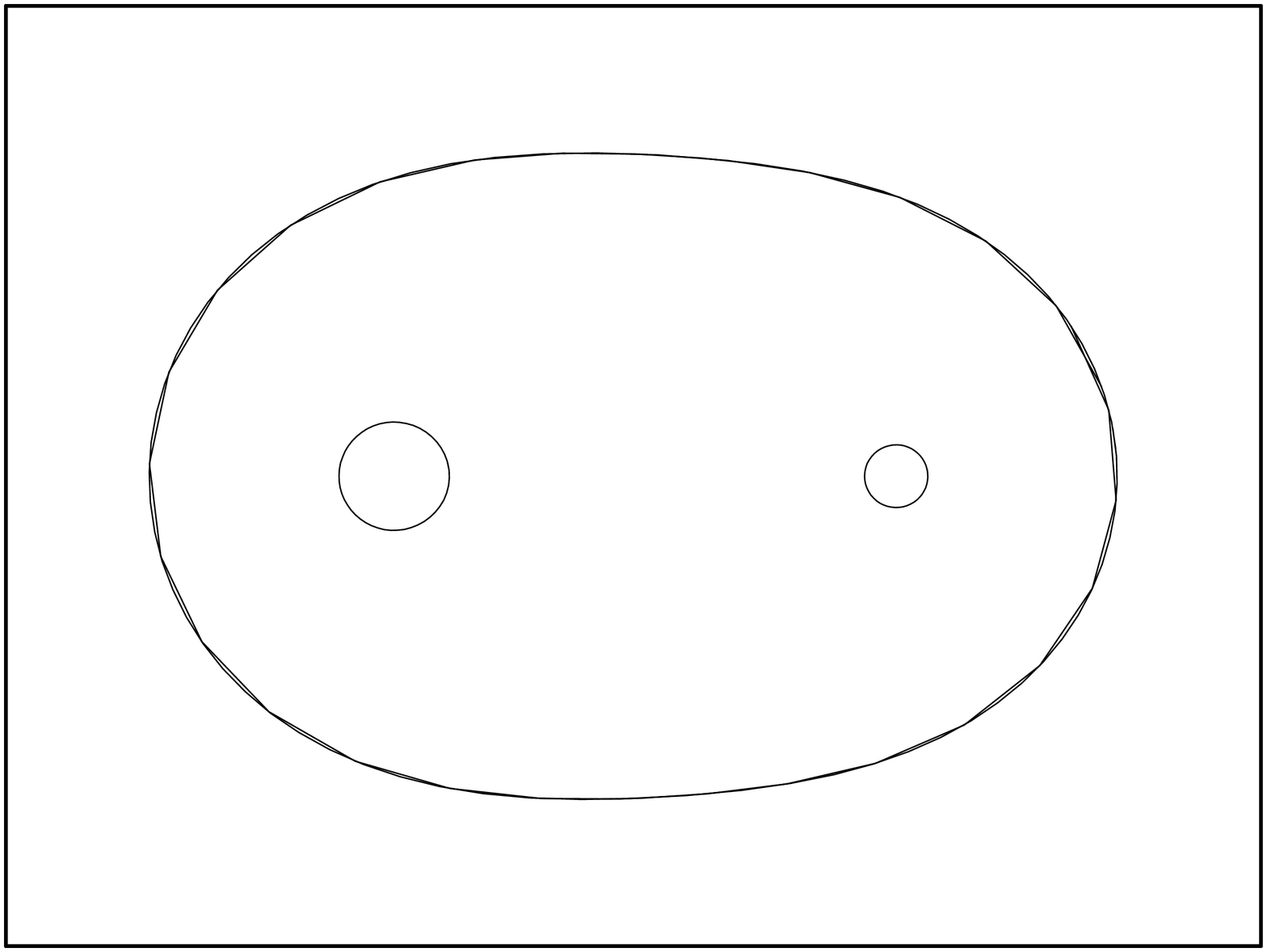, width=5cm, height=4.8cm }}
     \rotatebox{-90}{
    \epsfig{ figure = 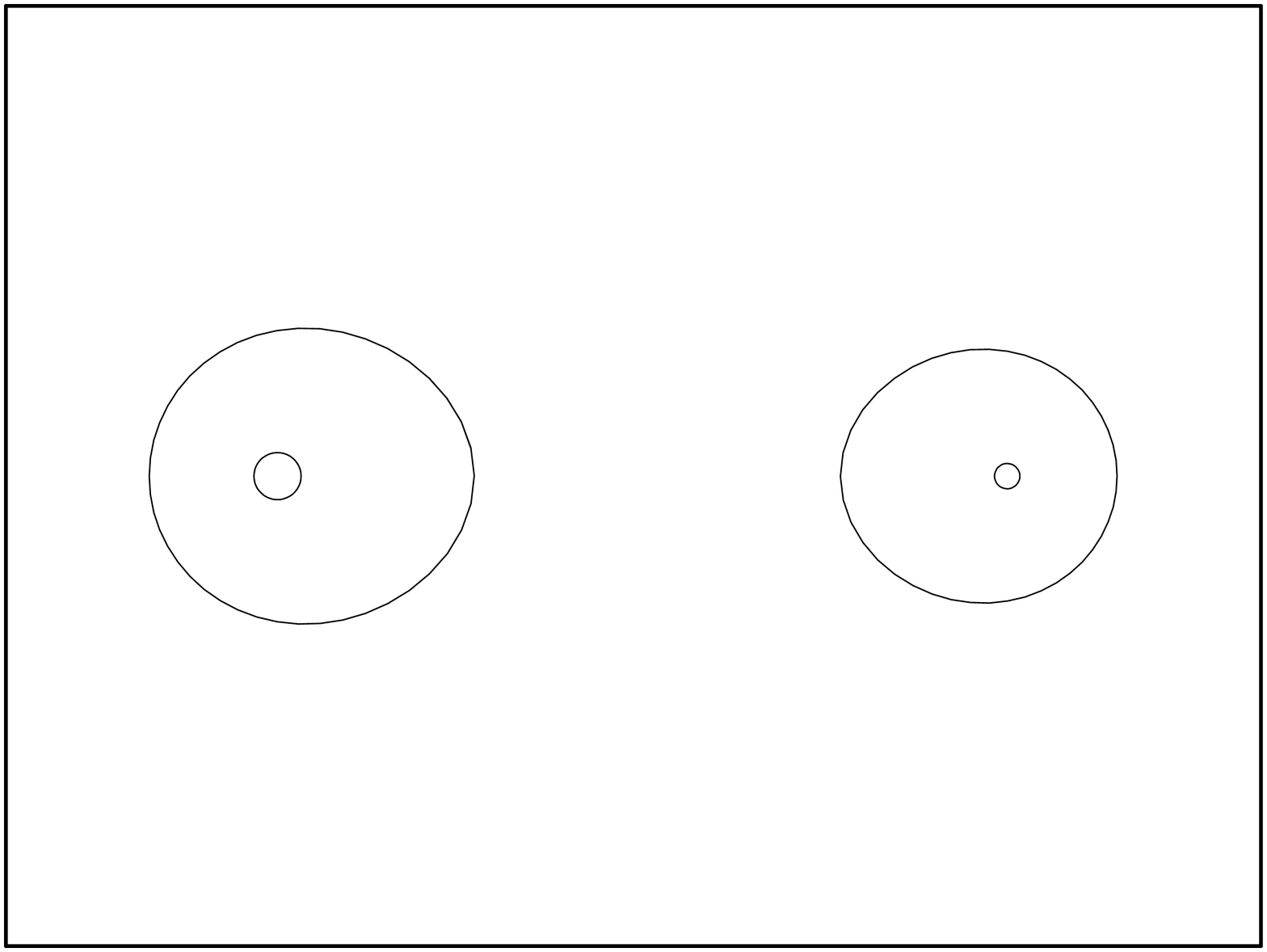, width=5cm, height=4.8cm }}
\caption{The inner and outer `horizons' in the \(xy\)-plane
      at fixed \(t\) for different \(x_{0}\).  }
 \label{fig:horizon}
\end{figure}

 To make the analogy with the BTZ case complete the
different regions we had can be generalized to,
\begin{equation}
\begin{array}{ll}
I:0<\rho<\alpha & \Rightarrow \ 0<\sqrt{\rho_1 \rho_2}<\alpha \\
II:\alpha<\rho<\alpha+\frac{\pi}{2} & \Rightarrow \
\alpha<\sqrt{\rho_1\rho_2} <\alpha+\frac{\pi}{2} \\
III:\alpha+\frac{\pi}{2}<\rho & \Rightarrow 
\alpha+\frac{\pi}{2}<\sqrt{\rho_1 \rho_2}
\end{array}
\end{equation}
In figure \ref{fig:horizon} we have plotted the `horizons' when we have
fixed \(r_+\) and \(r_-\) but varying distances \(x_0\) between the
singularities. Although the equations determining the boundaries of 
the regions are similar to the single-BTZ case we cannot be 
certain that we are dealing with true horizons, unless we trace light rays 
through the new geometries. This explains the quotation marks.

\subsection{Singularities}\label{subsec.DegMet}

The metric (\ref{SimpMultiBTZMet}) may locally be written 
\begin{equation}
    ds^2 = -g_{0}^2 dT^2 + g_{2}^2 d\Phi^2 +dh^2~,
    \label{eq:FormMet}
\end{equation}
with \(dT = r_{-}dt - f_{+}\) and \(d\Phi = r_{+}dt-f_{-}\), since \(
f_{+}\) and \( f _{-} \) are closed forms.  This metric degenerates
where \( g_{0} \) or \( g_{2} \) vanishes, where one of the functions
\( T(t,x,y) \), \(\Phi(t,x,y)\) or \( h(t,x,y) \) has critical point,
and where \( dT \), \(d\Phi \) and \( dh \) are linearly dependent. 
The coordinate singularities at the BTZ horizons and their multi-black
hole generalizations belong to the first case, but our solutions also
display the other types of degeneracies, and we now proceed to
investigate their interpretation.
 
In the case when one of the functions \( T \), \( \Phi \) or \( h \)
has a critical point, one may ignore the effects of the functions \(
g_{0} \) or \( g_{2} \) locally, since they may be absorbed into
redefinitions of \( T \), \( \Phi \) or \( h \) only in exceptional
cases at the expense of changing the nature of the critical point (but
see the following subsection to appreciate the importance of these
exceptions!).  Then the singularity is precisely of the kind discussed
by Horowitz \cite{Horowitz:1991qb} for zero cosmological constant. 
The simplest such singularity occurs between two equal charges
separated by some distance.

To see what happens we study the equal charge solutions close to the
origin.  There \(f_{+}=f_{-}=0\) because the contributions from the
two charges cancel by symmetry.  The metric (\ref{eq:FormMet}) then
degenerates at the origin at all times, because \( dT \) and \( d\Phi
\) both become parallel to the Killing direction \(dt\).  Furthermore,
\( h \), which approaches infinity at infinity and assumes local
minima at the positions of the charges, has to have a saddle point. 
Due to gauge invariance (\ref{eq:h-shift}) the position of the saddle
point may be chosen to be at the origin, making the metric on this
line (in spacetime) even more degenerate, of rank one.  Generically we
instead expect degenerations to rank-two metrics on two-dimensional
surfaces \cite{Horowitz:1991qb,Gib}.  In fact, we have found that the
map \( (t,x,y) \rightarrow (T,\Phi,h) \) has three singular fold
surfaces joined pairwise at three cusp lines if the saddle point of \(
h \) is displaced slightly.  The geometries of such complicated
singularities deserve a special study, but for our purposes it is
enough to find the simplest singularities in a gauge equivalence
class.

Returning to the case of coinciding saddles we proceed to determine 
the geometry close to the saddles. There we have approximately
\begin{equation}
    \begin{split}
        h=& a x^2 - b y^2\\ 
        r_{-} f_{+}=&r_{+} f_{-}= c \:d(x y) ~.
    \end{split}
    \label{eq:saddle}
\end{equation}
By rescaling coordinates and \(h\) we find a spatial line element
\begin{equation}
    ds^2 = d(xy)^2 + \frac{1}{4}d(x^2-y^2)^2 = 
    \left(x^2+y^2\right)\left(dx^2 + dy^2\right)
    \label{eq:metric}
\end{equation}
The area \(A_{O}\) and circumference \(C_{O}\) of circles around the origin 
are then related by \( C_{O}^2 = 8 \pi A_{O} \) in contrast to the 
Euclidean relation \( C^2 = 4 \pi A \). Since the metric is manifestly 
flat the difference can only be due to a conical singularity at the 
origin, and we conclude that there is a negative deficit angle of \( 
2 \pi \). 

We have argued that simple conical singularities with a surplus angle
of \( 2 \pi \) appear in the geometries with two equal sources
provided the gauge is chosen so that saddles of \(h \) coincide with
zeroes of \(f_{+}\) and \(f_{-}\).  For \(n\) sources \(h\) typically
has \(n-1\) saddles since it is chosen to have \( n \) local minima at
the sources and a maximum (infinity) at infinity.  Similarly \(f_{+}\)
and \(f_{-}\) typically have \(n-1\) zeroes, because of the \(n\)
sources and the behaviour at infinity.  If \(f_{+}\) and \(f_{-}\) are
proportional their zeroes coincide, and \(h\) may be chosen to have
saddles at the same points.  Fixing the behaviour of \(h\)
appropriately close to its saddles the local calculation is then the
same as between two sources, and we conclude that there are \(n-1\)
conical singularities.  Physically the proportionality of \(f_{+}\)
and \(f_{-}\) means that the sources all have the same ratio \(J/M\)
of spin and mass.  Other source distributions generally lead to more
complicated singularities in the geometry.  Some of these may be
removable like the coordinate singularites of the BTZ geometry, but
some are likely to be required by global arguments, like the conical
singularities we have just discussed.

\subsubsection{Geodesic singularities}\label{subsubsec.Geo}

The stationary conical singularities discussed above have been found
before by Clement \cite{Clement:1994qb} and by Coussaert and Henneaux
\cite{Coussaert:1994if}.  These authors have also remarked that such
singularities do not follow geodesics.  This is quite disturbing for
the commonly used hypothesis that Chern-Simons theory should be
relevant to the counting of black hole states.  Already at the
classical level would Chern-Simons theory give rise to geometries
which seem to leak energy and momentum!

Fortunately, Chern-Simons theory itself contains the answer to the
problem.  By asking under what precise conditions the singularities
are non-geodesic we may find an exception: when the singularity is
located at a horizon.  This case was already mentioned by Clement, but
not in the Chern-Simons context where it becomes truly important. 
While the set of geometries with singularities fixed to horizons may
seem like an exceptional set of measure zero, in Chern-Simons theory
they are not exceptional.  In fact, large class of solutions (and all
those considered by Coussaert and Henneaux) may be written in a gauge
such that the singularities are located at the horizon and thus follow
geodesics.  We now proceed to give some details of this argument.

We need to evaluate the Christoffel symbols \(\Gamma_{tt}^{x}\) and
\(\Gamma_{tt}^{y}\) which vanish precisely where there are static
geodesics.  If they can be made to vanish at the conical singularities
the puzzle of the unphysical geometries is solved.  We study the
Coussaert-Henneaux solutions, which are essentially ordinary BTZ
solutions, but with the mass and angular momentum distributed in the
same proportions on several sources.  In our language this means that
\begin{equation}
    r_{-}f_{+}=r_{+}f_{-}=r_{-}r_{+}f~,
    \label{eq:fdef}
\end{equation}
where the single form \(f\) encodes the source distribution. As has 
been pointed out several times above this assumption simplifies the 
interpretation of the solutions considerably. Now
\begin{equation}
\Gamma_{tt}^{x}=\pm( r_{+}^2-r_{-}^2) \frac{f_{x}
g_{i}(h){g_{i}}'(h)}
{f_{y}\partial_{x}h-f_{x}\partial_{y}h}~,
    \label{eq:Christoffel}
\end{equation}
where the sign (and the label \(i\)) depends on the region.  In
general this expression and the one for \(\Gamma_{tt}^{y}\) diverge
at a common zero of \( f \) and critical point of \( h \), but if \( g
\) or \( g' \) vanishes at the same point the whole expression instead
goes to zero.  This is what happens if the conical singularity is
located at a horizon.

It only remains to argue that the singularities can be moved to 
a horizon. Indeed, in region \(III\) a infinitesimal shift of \(h\) is 
equivalent to an infinitesimal gauge transformation 
(\ref{eq:h-shift}), and similar relations exist in the other regions. 
Assuming that these transformations can be integrated, we conclude 
that changes of function \(h\) are gauge transformations. By adjusting 
\(h\) we can then make \( g \) or \( g' \) vanish at a conical 
singularity, i.e. a gauge transformation may take the singularity to 
a horizon, where it follows a (null) geodesic simply by being stationary.

\section{Conclusions}\label{sect:Conc}

We have constructed and investigated solutions to three-dimensional
AdS gravity which generalize the BTZ solution.  While the ordinary BTZ
black hole can be viewed as a single source solution in the
Chern-Simons formulation, we have constructed multi-source solutions. 
These solutions give rise to a kind of multi-black hole solutions,
which however also display other singularities.  In the simplest cases
the additional singularities are fixed conical singularities, but more
complicated cases also occur.  Einstein's equations break down at
these singularities, so they represent geometries which are not
allowed in pure einsteinian gravity.  On the other hand, they occur
very naturally in the Chern-Simons framework, which is natural for
quantization, so we believe that these multi-black hole solutions
should be included in a full Chern-Simons treatment of BTZ black hole
entropy.

We have also shown that a large class of these multi-black hole
solutions allow a gauge choice which ensures that the singularities in
the corresponding geometries follow geodesics.  Geometrically the
solutions then precisely encode the BTZ solution outside a number of
horizons.  These horizons are however all connected with each other,
since the conical singularities which join them can not appear outside
the horizons without violating the geodesic equation.  The union of
all these horizons appears to the outside observer as a single
horizon.  Only at the horizon (and inside) is the difference to the
single black hole solution noticable.  In this picture of a single
horizon, special light-like geodesics on the horizon are 
identified pairwise, since they in fact represent the same conical 
singularity, only approached from two different directions (two 
different ridges on the saddle point of the function \(h\)).

Although we have not attempted in this paper to find the quantum 
states corresponding to the multi-black hole solutions, we have 
provided evidence that such states should be included in the black 
hole spectrum. Namely, the asymptotics at infinity of the classical 
solutions approach the single-BTZ solutions so rapidly that the 
difference can not be detected by any asymptotic charges. Only 
non-asymptotic observables like the holonomies distinguish between 
the solutions. It then seems quite unnatural to exclude the sectors 
with multiple sources, in particular since the sources may be hidden 
inside the horizon. Presumably, the additional sectors of the boundary 
conformal field theory that are required to represent multi-black hole 
solutions can also be understood by purely two-dimensional 
considerations, for instance by the requirement of modular invariance.

\section{Acknowledgements}

We would like to thank S\"oren Holst and Max Karlovini for useful
discussions.  It is a pleasure to also thank Marc Henneaux for a
conversation about reference \cite{Coussaert:1994if} and Ingemar
Bengtsson for one on papers \cite{Aminneborg:1998pz,Aminneborg:1999si}. 
The work of B. S. was financed by the Swedish Science Research
Council.

\end{document}